\documentclass[12pt]{article}
\usepackage{amssymb,theorem,amsmath}
\usepackage{enumerate}

\usepackage{graphicx}
\newcommand{\Var}{\mathop{\rm Var}}

\def\csect#1{Section~\ref{#1}}

\def\csectt#1#2{Sections~\ref{#1}--\ref{#2}}
\let\Prp=\Pr \def\Pr{\Prp\nolimits}
\newcommand{\be}{\begin{equation}}
\newcommand{\ee}{\end{equation}}

\newtheorem{theorem}{Theorem}%[]

{\theorembodyfont{\rm} 
\newtheorem{remark}{Remark}

 }

\newcommand{\cthm}[1]{Theorem~\ref{#1}}

\newcommand{\cfig}[1]{Figure~\ref{#1}}

\newcommand{\crem}[1]{Remark~\ref{#1}}

    \def\bbz{\mathbb{Z}}

    \def\bbr{\mathbb{R}}

\def\ss{\scriptstyle}\def\sss{\scriptscriptstyle}
\def\interiorM{\rlap{\raise9pt\hbox to9.5pt{\hss$\sss\circ$}}M}
\long\def\kill#1\endkill{\relax}

\def\0{{\it0}}\def\1{{\it1}}\def\2{{\it2}}\def\3{{\it3}}

\def\Fe{F_{\rm even}}\def\Fo{F_{\rm odd}}

\def\bll{\noindent\hbox to 15pt{\hfil$\bullet$\hfil}}

\begin{document} \title{Approach to Hyperuniformity of Steady States
of Facilitated Exchange Processes}

\author{S. Goldstein\footnote{Department of Mathematics,
Rutgers University, New Brunswick, NJ 08903.},
J. L. Lebowitz\footnotemark[1],
\footnote{Also Department of Physics, Rutgers.}\ \ 
and E. R. Speer\footnotemark[1]}
\date{January 29, 2024}

\maketitle

\begin{flushleft}\noindent {\bf Keywords:} Facilitated exclusion
  process, conserved lattice gas, hyperuniform states.
\end{flushleft}

\begin{abstract} We consider the fluctuations in the number of
particles in a box of size $L^d$ in $\bbz^d$, $d\ge1$, in the
(infinite volume) translation invariant stationary states of the
facilitated exclusion process, also called the conserved lattice gas
model.  When started in a Bernoulli (product) measure at density
$\rho$, these systems approach, as $t\to\infty$, a ``frozen'' state
for $\rho\le\rho_c$, with $\rho_c=1/2$ for $d=1$ and $\rho_c<1/2$ for
$d\ge2$. At $\rho=\rho_c$ the limiting state is hyperuniform, that is,
the variance of the number of particles in the box grows slower than
$L^d$.  We give a general description of how the variances at
different scales of $L$ behave as $\rho\nearrow\rho_c$. On the largest
scale, $L\gg L_2$, the fluctuations are normal (in fact the same as in
the original product measure), while in a region $L_1\ll L\ll L_2$,
with both $L_1$ and $L_2$ going to infinity as $\rho\nearrow\rho_c$,
the variance grows faster than normal.  For $1\ll L\ll L_1$ the
variance is the same as in the hyperuniform system.  (All results
discussed are rigorous for $d=1$ and based on simulations for
$d\ge2$.)  \end{abstract}

\section{Introduction\label{intro}}

   An important quantity in the study of particle systems in $\bbz^d$
(or $\bbr^d$) is the fluctuation in the number of particles in a
domain $D$; an indicator of the size of such fluctuations is the
variance of the number of particles in $D$. We will restrict our
attention to particle systems on $\bbz^d$ in which each lattice
site can contain at most one particle.   Consider then a
translation invariant system in which the probabilities of
configurations are described by a measure $\mu$ and the particle
density is $\rho$, and let $N_L$ be the number of particles in a
cubical domain of size $L^d$. The expected value of $N_L$ is
$\rho L^d$ for all $L$, and we shall denote its variance by $V(L)$.
In general $V(L)$ is related to the pair correlation function:
$V(L)/L^d\to S(0)$ as $L\to\infty$, where $S(k)$ is the Fourier
transform of the total pair correlation function \cite{GL}.

For many systems of this sort, such as equilibrium systems which are
described by a Gibbs measure with integrable potentials and are not at
a phase transition, $V(L)\simeq CL^d$ as $L\to\infty$, with $C$ the
compressibility.  For an ideal lattice gas (i.e., one with no
interactions between different lattice sites) in equilibrium, the
Gibbs measure $\mu$ is a Bernoulli product measure and
$C=\rho(1-\rho)$.  There are also many cases of interest, such as
systems at equilibrium critical points or the voter model in $d>2$
(for which the result follows from (2.4) of \cite{ps}), in which
$V(L)$ grows faster than $L^d$ and $S(0)$ is infinite.

Our interest here is in the opposite case, in which $V(L)$ grows more
slowly than the volume, i.e., $S(0)=0$.  These are the so called {\it
hyperuniform} systems \cite{GL,T} (originally called super homogeneous
systems).  In fact we shall restrict our considerations to the
particular situation in which the measures studied describe the final
stationary states of stochastic lattice systems started from a
Bernoulli product measure and evolving according to the {\it
facilitated exclusion process} (FEP) (also called the {\it conserved
lattice gas model}). As we shall describe in detail below, these
systems undergo a transition from frozen to active final states at
some critical density $\rho_c$, at which density the system is
hyperuniform.

The approach to this hyperuniform state as $\rho\to\rho_c$ depends on
the direction of approach and is very different, for
$\rho\nearrow\rho_c$, from what one might expect in general, which is
that the constant $C=C(\rho)$ would approach zero (as in the ideal
lattice gas when $\rho\to1$).  Rather, when the approach is from lower
densities, $C(\rho)$ continues to have its ideal gas value for all
$\rho$, but the values of $L$ at which the approximation
$V(L)\approx C(\rho)L^d$ is valid become larger and larger as
$\rho\nearrow\rho_c$.  For approach from above we do have
$C(\rho)\searrow0$ as $\rho\searrow\rho_c$.  In particular for $d=1$,
where $\rho_c=1/2$, it follows from (3.3) of \cite{del} that
$C(\rho)=\rho(1-\rho)(2\rho-1)$, since the stationary measure for
``extended particle'' model studied there, with particles covering $2$
sites, is shown in \cite{agls} to coincide with the stationary measure
for the FEP at densities $\rho>1/2$ after interchange of particles and
holes.

In the remainder of the paper we first (\csect{facsys}) introduce the
facilitated exclusion systems that we study and review earlier work,
in particular the work by Hexner and Levine \cite{hl} discussing, for
the two-dimensional model, the hyperuniformity of the system at the
critical density and the approach to this hyperuniformity as
$\rho\nearrow\rho_c$.  We then, in \csect{behavior}, propose a more
detailed, but still phenomenological, characterization of the
$L$-dependence of the variances $V(L)$ for $d=1,2$, supporting the
proposal by a procedure we call {\it power scaling}.  We also present
a simple argument, rigorous for $d=1$, for the fact that in all these
models, and at all densities $\rho<\rho_c$, the variance grows as
$\rho(1-\rho)L^d$ for sufficiently large $L$ (that is, the prefactor
$C$ mentioned above is then $\rho(1-\rho)$).

The proposal in \csect{behavior} is based on data from simulations,
some new, in two dimensions; in \csect{model} we describe in more
detail the FEP process in $d=1$.  In that case we can establish
rigorously a version of the general structure described in $d=2$.  We
give exact values for the exponents characterizing the behavior of the
variance in different ranges of $L$; these are different from those in
$d=2$, and in $d=1$ there is also a difference, for $L$ small, between
the exponents for $L$ odd and for $L$ even.  In \csect{details} we
give a hint of the rigorous derivation of these results; full proofs
will be given elsewhere \cite{GLSrigor}.  In the concluding section we
mention results for some other related models and discuss open
problems.

\section{Facilitated systems\label{facsys}} 

We shall be concerned here primarily with certain stationary states,
in $\bbz^d$, of the symmetric Facilitated Exclusion Process (FEP),
also known as the Conserved Lattice Gas process.  In this system
a site of the lattice can be occupied by at most one particle, and a
particle can jump to a neighboring empty site only if it has also an
occupied neighbor site.

The stationary states of this system have been investigated
numerically for $d\ge2$ \cite{hl,rpv,lub} and theoretically for $d=1$
\cite{agls,BESS,BES,glsshort,GLS2}.  These investigations suggest that the
system, when started with an initial random configuration at density
$\rho$, approaches, as $t\to\infty$, either a {\it frozen} state in
which all particles are isolated and hence unable to move, or an {\it
active} stationary state in which there is a finite density of {\it
active} particles: those with an occupied neighboring site.  The
transition (often called a {\it transition to an absorbing state} or
an {\it absorbing phase transition}) occurs at a density
$\rho_c\le1/2$ (for $\rho>1/2$ it is geometrically impossible for all
the particles to be isolated).  For $d=1$, $\rho_c=1/2$, but rather
surprisingly, when $d\ge2$ the simulations find values of $\rho_c$ to
be much smaller than $1/2$. We discuss below the determination of
$\rho_c$ as well as its dependence on which of several different ways
the dynamics of the model is defined.

Another surprising observation, due to Hexner and Levine \cite{hl}, is
that, in 2d and 3d, the frozen limiting state of the system at
$\rho_c$ is hyperuniform: $V(L)\sim L^{\lambda_1}$, with
$\lambda_1\approx1.57$ in 2d and $\lambda_1\approx2.76$ in 3d
\cite{hl}.  (Throughout we will relate two quantities by ``$\simeq$''
or, respectively, by ``$\sim$'' to express the fact that asymptotically
their ratio is equal to 1 or, respectively, is bounded away from both
0 and $\infty$.)  Hexner and Levine also discuss in detail how $V(L)$
grows with $L$ for $\rho$ close to, but less than, $\rho_c$.  They
find a crossover, as $L$ increases, from a scale on which $V(L)$ grows
as $L^{\lambda_1}$ to a scale on which the growth is as $L^d$ (in each
case for $L$ large on the relevant scale).  In the current paper we
are particularly interested in the transition region between the two
behaviors, the location of which moves towards larger values of $L$ as
$\rho\nearrow\rho_c$.

 \medskip\noindent
 {\bf Dynamics and the determination of $\rho_c$:} Several different
implementations of the dynamics have appeared in the literature.  The
jumps may occur either in continuous time, with each particle
attempting to jump at rate 1, or in discrete time, with all active
particles attempting to jump simultaneously at integer times; the
target site for the jump can be (1)~chosen at random from all
neighbors, with a jump occurring only if the chosen site is empty, or
(2)~chosen at random from just the empty neighbors.  (In the discrete
time case several particles may attempt to jump to the same site; in
this case, one is chosen at random to succeed.)  For example, discrete
time updating with rule (2) is used in \cite{rpv,hl}; continuous time
updating is used with rule (2) in \cite{lub} and with rule (1) both in
\cite{oliveira} and, for {\it exclusion processes} (in which no
facilitation is required for a jump), in the mathematical literature
of interacting particle systems \cite{Liggett}.

 To investigate $\rho_c$ for $d=2$ one simulates the dynamics in a
square box in $\bbz^2$, with periodic boundary conditions, starting
with a random distribution at some density $\rho<1/2$.  In this
setting it is easy to show, using the theory of finite state Markov
processes, that for any initial state the system {\it must} eventually
freeze.  This means that a non-frozen steady state with $\rho<1/2$
cannot exist in finite volume, and hence that if one were to try to
define a critical density for a finite system as we have done for the
infinite system, one would find a value of $1/2$.  To avoid mentioning
this uninteresting result further  we reserve the term ``critical density''
and the notation $\rho_c$ for the infinite-system value.

Nevertheless, there are several methods for determining an approximate
value of $\rho_c$ from such simulations; we discuss two of these.  The
first method is straightforward, depending only on the observation of
dramatic growth in the freezing time for $\rho$ near $\rho_c$.  The
second method requires making some assumptions about the behavior of
the infinite system as $\rho\searrow\rho_c$.  It gives precise values
for $\rho_c$ (which depend, however, on the size of the simulated
system).  For the sake of definiteness we will adopt here the value
which we obtain from this method in a box of side $L=1024$.

\begin{figure}[ht!]\begin{center} \hbox to
\hsize{\hss\includegraphics[scale=0.70]{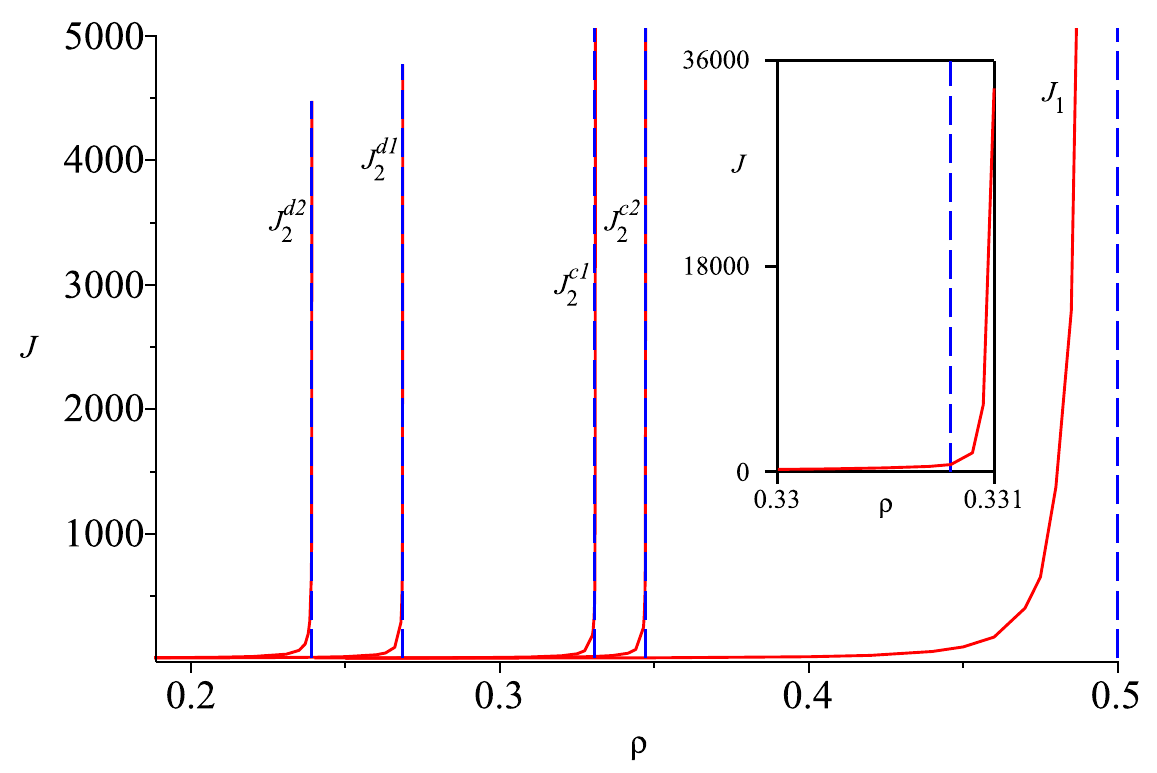}}

\caption{Mean value of the number $J$ of jumps per particle required to
freeze the system, versus density $\rho$, for the 2d FEP with four
different dynamics; $J^{d2}_2$ denotes the curve for the
distrete-time rule (2) model, etc.  Also shown, and denoted $J_1$,
are results for the 1d continuous-time FEP. The (dashed) vertical
lines are at the critical densities.  The inset shows an expanded view
of the curve $J_2^{c1}$ near its critical value $\rho_c=0.3308$.} \label{fig:Trho}
\end{center}\end{figure}

The first approach is to count the number of particle jumps required
for the system to freeze; for a somewhat vaguely defined density
$\tilde\rho_c$ this number increases very rapidly as
$\rho\nearrow\tilde\rho_c$.  Let $J=J(\rho)$ be the number of required
jumps divided by the number of particles in the system.
\cfig{fig:Trho} plots the mean of $J(\rho)$, over 200 trials, against $\rho$,
for the four types of dynamics discussed above; the data are for a box
of side $L=1024$.  (We have also shown in this figure the values of
$\rho_c$; clearly one could take $\tilde\rho_c=\rho_c$, since the
sharp increase in $J(\rho)$ at $\rho_c$ is clear.)  For comparison the
results of a 1d simulation (in a periodic system of $2^{17}$ sites) of
the continuous-time FEP, for which $\rho_c$ is known to have value
$1/2$, are also shown.

In fact the situation is not quite as simple as suggested above and by
the main part of \cfig{fig:Trho}: for the 2d models there {\it is} a
sharp increase in $J$ as the critical density is approached, but a
much more rapid one just past that density.  This is shown, for the
continuous-time rule (1) model, in the inset.  In contrast, for the 1d
model the increase occurs well before the critical value.

The second method \cite{lub} is first to determine an approximate
value of $\rho_c$, perhaps by the method discussed above, and then to
simulate the system at several slightly higher densities.  At such
densities the simulation will enter a long-lived metastable state
which is presumably close to a projection of the infinite-volume
active state at that density.  Let $\rho_a=\rho_a(\rho)$ denote the
density in this state of active particles; then $\rho_a$ vanishes as
$\rho\searrow\rho_c$ and one hypothesizes that it does so with a power
law: $\rho_a\simeq C(\rho-\rho_c)^\beta$.  $\rho_c$, $C$, and $\beta$
are then taken to be the parameters which give the best fit to this
relation, using a least-squares fit for the linear relation between
$\log\rho_a$ and $\log(\rho-\rho_c)$.  With this method we find, from
simulations in an $L\times L$ box with $L=1024$, that (i) for the
continuous time models, $\rho_c=0.3308$ for dynamics defined by rule
(1) and $\rho_c=0.3471$ for rule (2), and (ii) for the discrete time
models, $\rho_c=0.2685$ for rule (1) and $\rho_c=0.2391$ for rule (2).
For all four models the values of $\beta$ lie in the interval
$[0.621,0.628]$.

\section{Phenomenological description\label{behavior}}

As noted in the introduction, we wish to study the model in the region
$\rho\le\rho_c$; from now on we restrict our attention to this region
and write $\delta:=\rho_c-\rho\ge0$.  Let $\mu_t^{(\rho)}$ be the
infinite-lattice measure at time $t$ when the initial measure is the
Bernoulli (product) measure of density $\rho$, which we denote
$\mu_0^{(\rho)}$, and let
 \be\label{nurho}
  \nu_\rho:=\lim_{t\to\infty}\mu_t^{(\rho)}
 \ee
  be the corresponding final state.  We will always assume that
$\nu_\rho$ is frozen; we know of one model for which this is not true
when $\rho=\rho_c$, the 1-d discrete-time symmetric FEP \cite{GLS3},
and will describe this in \csect{conclusion} but not consider it here.

  The following picture of the behavior of the variances $V(L)$, under
the measure $\nu_\rho$, as $\rho\nearrow\rho_c$, emerges from
simulations of the FEP in two dimensions and theoretical
considerations in $d=1$; we discuss it, however, in an arbitrary
dimension $d$.  Suppose that $\rho<\rho_c$ and that $\delta\ll1$.
Then:

 \begin{enumerate}[(P1)]

\item There are three regimes in $L$, which we call regimes of ``small,''
``intermediate,'' and ``large'' $L$, although we emphasize that since we
are speaking of asymptotic results, our descriptions always assume
that $L$ is in fact sufficiently large.

\item In the regime of {\it small} $L$ the variances grow
approximately as in the hyperuniform state at $\rho_c$:
$V(L)\simeq C_1L^{\lambda_1}$ (see \csect{facsys}).

 \item At some (approximately defined) scale $L_1=L_1(\delta)$ the
variances enter the regime of {\it intermediate} $L$, in which they
grow as $V(L)\simeq C_2(\delta)L^{\lambda_2}$ with $\lambda_2>d>\lambda_1$
and $C_2(\delta)>0$.

\item The growing variances reach size of order $L^d$ at an
(approximate) scale $L_2=L_2(\delta)$; for $L>L_2$, the regime of {\it
} L, the variances grow so as to coincide, even up to prefactor,
with the variances in the initial measure: $V(L)\simeq\rho(1-\rho)L^d$.

\item As $\rho\nearrow\rho_c$, $L_1(\delta)$ and $L_2(\delta)$
increase as $L_i\sim
\delta^{-\gamma_i}$ for some exponents $\gamma_1,\gamma_2$, with
$\gamma_2>\gamma_1>0$.  \end{enumerate}

\begin{figure}[ht!]\begin{center} \hbox to
\hsize{\hss\includegraphics[scale=0.70]{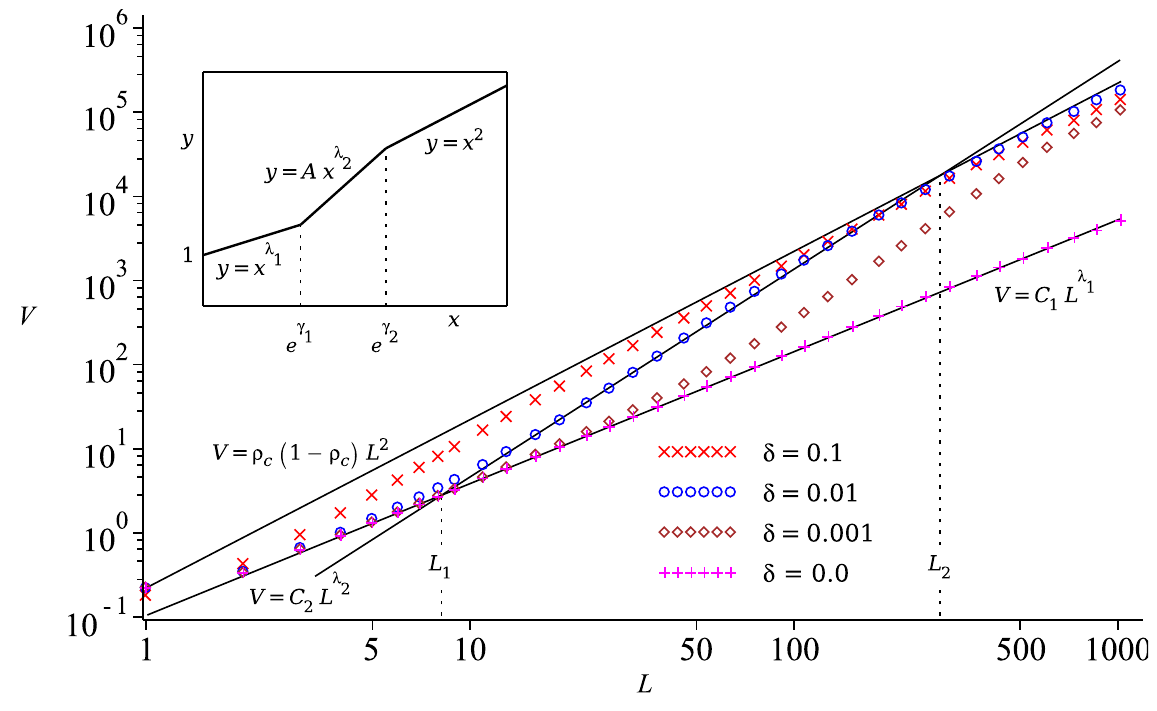}}

\caption{Log-log plot of the variances $V(L)$ in the continuous-time
rule-(1) 2d FEP for four density values, together with the
approximating straight lines giving the power law growth described in
(P1)--(P5).  The values of $L_1$, $L_2$, and $C_2$ shown correspond to
$\delta=0.01$. The inset shows the conjectured limit of the data under
power scaling; if $C_2(\delta)=D\delta^\theta$ as in \crem{gammas} then
$A=e^{-\theta}$.} \label{fig:essence} \end{center}\end{figure}

We illustrate (P1)---(P5) in \cfig{fig:essence}, a log-log plot of the
variances $V(L)$ in the continuous-time, rule (1), 2d FEP for
$\delta=0.1$, 0.01, 0.001, and 0.  The data is from a $2048\times2048$
system.  Also shown are the lines $V=C_1L^{\lambda_1}$ (here
$C_1=0.104$ and $\lambda_1=1.57$), $V=\rho_c(1-\rho_c)L^2$, and, for
$\delta=0.01$, the line $C_2(\delta)L^{\lambda_2}$ (with
$C_2(0.01)=0.0157$, $\lambda_2=2.47$) and approximate values of
$L_1=L_1(0.01)$ and $L_2=L_2(0.01)$.  The inset will be discussed
below.

 \medskip\noindent
 {\bf Power scaling:} We can further justify the conclusions
(P1)---(P5) by collapsing the data under an appropriate scaling.  In
\cite{hl} a multiplicative scaling, which in our terms would be
$L\to\delta^{\gamma_1}L$, $V\to\delta^{\lambda_1\gamma_1}V$, partially
collapses the observed $L$-$V$ data for the 2d FEP.  As remarked
there, however, such a scaling cannot give complete collapse, since
there are different exponents in the different regimes.  Instead we
introduce a {\it power scaling}:
 \be\label{scaling}
\bigl(L,V(L)\bigr)
   \to\bigl(L^{\alpha(\delta)},V(L)^{\alpha(\delta)}\bigr),
 \ee
 which corresponds to a {\it linear} rescaling in the log-log plots,
carrying straight lines to straight lines of the same slope.  The
exponent $\alpha(\delta)$ must be chosen so that the scaling limit
exists when $\delta\to0$; we choose
$\alpha(\delta)=-(\log\delta)^{-1}$, which leads to limits for the two
``special'' points, $(L_1,C_1L_1^{\lambda_1})$ and $(L_2,L_2^d)$, of
\cfig{fig:essence}(a):
\begin{align}\nonumber
  \bigl(L_1,C_1L_1^{\lambda_1}\bigr)
   \sim(\delta^{-\gamma_1},C_1\delta^{-\gamma_1\lambda_1})
 &\rlap{\hbox to 34pt{\hss$\longrightarrow$\hss}}
 \raisebox{-1.8ex}{\hbox to 34pt{\hss$\ss\rm scale$\hss}}
  \bigl(e^{\gamma_1},
     C_1^{\alpha(\delta)}e^{\gamma_1\lambda_1}\bigr)\\
  & \rlap{\hbox to 34pt{\hss$\longrightarrow$\hss}}
 \raisebox{-1.8ex}{\hbox to 34pt{\hss$\ss\delta\searrow0$\hss}}
  \bigl(e^{\gamma_1},
     e^{\gamma_1\lambda_1}\bigr),\label{sclim}
\end{align}
  and similarly, under the same scaling and limit,
$(L_2,L_2^d)\to(e^{\gamma_2}, e^{\gamma_2d})$. Thus under this scaling
the plot of \cfig{fig:essence} approaches the limit shown in the inset
of that figure.  

\begin{figure}[ht!]\begin{center} \hbox to
\hsize{\hss\includegraphics[scale=0.70]{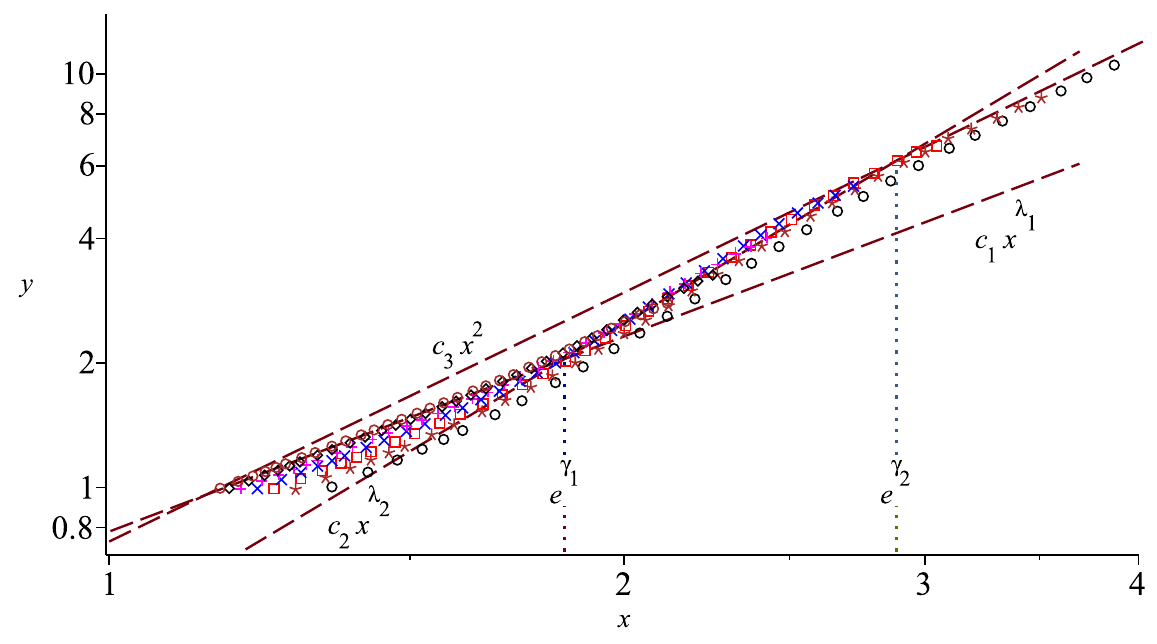}}

\caption{Log-log plot of the variances $V(L)$ in the continuous-time
rule-(1) 2d FEP after power scaling, for density values $\delta=0.01$,
0.004, 0.002, 0.001, 0.0004, 0.0002, and 0.0001. Also shown are
approximating straight lines giving the power law growth and the
approximate values $e^{\gamma_1}$ and $e^{\gamma_2}$, which here
correspond to $\gamma_1\approx0.61$, $\gamma_2\approx1.06$.}
\label{fig:2dscaled} \end{center}\end{figure}

The effect of power scaling for actual data---specifically, that of
the con\- tinuous-time, rule (1) model at seven different densities,
from $\delta=0.01$ to $\delta=0.0001$---is shown in
\cfig{fig:2dscaled}.  Also shown are approximate values of
$e^{\gamma_1}$ and $e^{\gamma_2}$ obtained from \eqref{sclim}.  The
constants $c_1$, $c_2$, and $c_3$ used in the figure for the curves
$y=c_ix^{\lambda_i}$ are chosen to fit the data shown; in the
$\delta\searrow0$ limit $c_1$ and $c_3$ would have value 1, but the
logarithmic convergence of $\alpha(\delta)$ to zero with $\delta$ is
too slow for this limit to be apparent.

 \medskip\noindent
 {\bf Further discussion:} The existence of a regime of small $L$ with
behavior as described in (P2) can be readily understood, given that
hyperuniformity holds for $\nu_{\rho_c}$ as described in
\csect{facsys}.  For as $\delta\searrow0$ the state $\nu_\rho$ should
converge to $\nu_{\rho_c}$, which implies that the distribution under
$\nu_\rho$ of the number of particles in the box $B_L$, and hence in
particular the variance $V(L)$ of that number, should converge to that
for $\nu_{\rho_c}$. (Of course for larger and larger $L$ we must
require smaller and smaller $\delta$.)  Thus for $0<\delta\ll1$ we
should have that $V(L)\simeq C_1L^{\lambda_1}$, provided that $L$ is
large but not too large.

The fact that the fluctuations are as in the initial measure for
sufficiently large $L$, that is, that as described in (P4),
$V(L)\simeq \rho(1-\rho)L^d$ there, is also easily understood.  For,
since the configurations are becoming frozen as $t\nearrow\infty$, we
expect that each particle will, with probability one, move only a
finite distance during the evolution.  Thus the number of particles in
a sufficiently large box is, to high relative accuracy, the same at
the end of the evolution as it was at the beginning
\cite{glsshort,GLS2}, and the fluctuations in this number should also
be the same as for the original Bernoulli measure.

There must thus be a region of transition between the region of small
$L$ with slow growth $V(L)\simeq C_1L^{\lambda_1}$ and the region of
large $L$ with normal $\rho(1-\rho)L^d$ growth; the growth in this
intermediate region must be faster than $V(L)\simeq \rho(1-\rho)L^d$ to
connect the smallish slow-growth variance for small $L$ with the
normal growth variance at large $L$.  One possibility is that of (P3),
$V(L)\simeq C_2(\delta)L^{\lambda_2}$ with $\lambda_2>\lambda_1$ and
$\lambda_2>d$, but we see no {\it a priori} reason for this to hold;
one might also have, for example, normal growth $V(L)\simeq CL^d$ with
prefactor $C$ larger than $\rho_c(1-\rho_c)$.

\begin{remark}\label{gammas}Suppose that in fact
$V\simeq C_2(\delta)L^{\lambda_2}$ in the intermediate region, as in
(P3), and that for some constant $D$, $C_2(\delta)\simeq D\delta^\theta$
as $\delta\searrow0$. (This is in fact true for the $d=1$ case, with
$\theta=1$ and $D=\sqrt{8/\pi}/3$; see \csect{model}).  This will then
determine the exponents $\gamma_1$ and $\gamma_2$ of (P5): matching
values of $V(L)$ at $L_1(\delta)$ and $L_2(\delta)$ yields
$C_1L_1^{\lambda_1}=D\delta^\theta L_1^{\lambda_2}$ and
$\rho_c(1-\rho_c)L_2^d=D\delta^\theta L_2^{\lambda_2}$, from which we
obtain that $L_i(\delta)\simeq B_i\delta^{-\gamma_i}$ with
 \be\label{gammavals}
\gamma_1=\theta/(\lambda_2-\lambda_1), \qquad
\gamma_2=\theta/(\lambda_2-d),
 \ee
 and 
 \be\label{Bvals}
 B_1=(C_1/D)^{1/(\lambda_2-\lambda_1)},\qquad
  B_2(\rho_c(1-\rho_c)/D)^{1/(\lambda_2-d)}.
 \ee
 However, the status of $B_1$ and $B_2$ is somewhat different from
that of $\gamma_1$ and $\gamma_2$, as we discuss in \crem{more} in
\csect{model}.  \end{remark}

\section{The FEP in one dimension\label{model}}

We now turn to the FEP in one dimension, that is, on $\bbz$, still in
continuous time.  Note that in this case a particle can jump only if
exactly one of its two neighboring sites is occupied, so that at most
one jump, either left or right, is possible for each active particle
(so there is now no difference between the two dynamical rules (1) and
(2) of \csect{facsys}). .  For this model we can establish rigorously
\cite{GLSrigor} a slightly modified version of the properties
(P1)--(P5), as we now discuss; the proof also determines explicitly
the various parameters $\lambda_1$, $\lambda_2$, $C_1$, $C_2$,
$\delta_1$, and $\delta_2$ which appear in (P2), (P3), and (P5) and
confirms the behavior $V_\rho(L)\simeq\rho(1-\rho)L$ in the region of
``large'' $L$ (see (P4)).

In this section we write $V_\rho(L)$ for the variance of the number of
particles in an interval of length $L$ under the measure $\nu_\rho$ of
\eqref{nurho}.  The key input for our analysis of this quantity is the
explicit knowledge \cite{glsshort,GLS2} of $\nu_\rho$,
$\rho\le\rho_c=1/2$, for this model.  As a consequence our results
also hold for several other models, for which the measure $\nu_\rho$
agrees with that in the 1d FEP, as we discuss in \csect{conclusion}.
The measure $\nu_{1/2}$ at the critical density is of special
interest:
 \be\label{muhalf}
\nu_{1/2}
   =\frac12(\delta_{\eta^*}+\delta_{\eta^\dagger}),
 \ee
 where $\eta^*$ and $\eta^\dagger$ are the two configurations in which
holes and particles strictly alternate.  $\nu_{1/2}$ is (trivially)
hyperuniform; in particular,
 \be 
 V_{1/2}(L)=\begin{cases}\frac14,& \text{if $L$ is odd,}\\
   0,& \text{if $L$ is even.}\label{v2l}\end{cases}
 \ee

\begin{figure}[ht!]\begin{center} \hbox to
\hsize{\hss\includegraphics[scale=0.70]{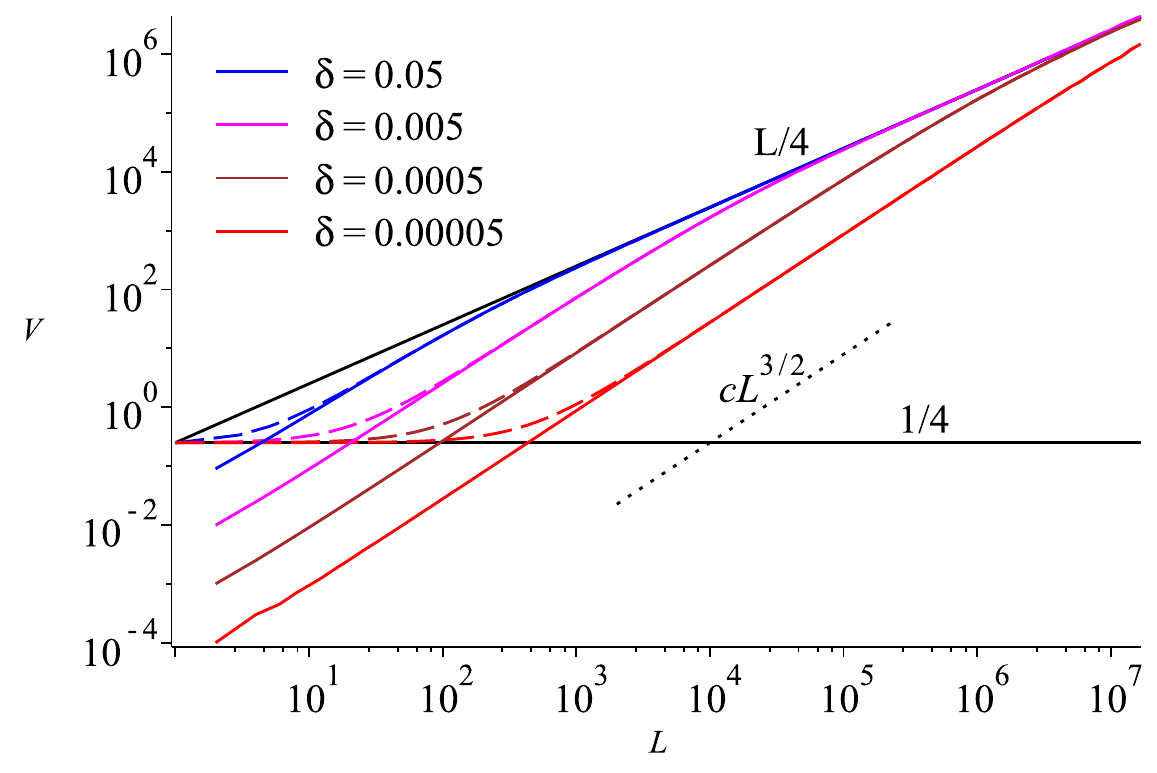}\hss}
\caption{Variances in the 1d FEP for four
values of the density.  The solid curves correspond to even values of $L$,
the dashed ones to odd values.  The black lines are for comparison
purposes: $V=1$, $V=L/4$, and $V=cL^{3/2}$ (dotted).} \label{fig:unscaled}
\end{center}\end{figure}

  Plots of $V_\rho(L)$ for four values of $\rho$ are shown in
\cfig{fig:unscaled}.  (The data
plotted in \cfig{fig:unscaled} were obtained by direct simulation of
the known final measure $\nu_\rho$, without actually simulating the
time evolution; this method permits the evaluation of $V_\rho(L)$ for
large values of $L$.) It is clear that the behavior of $V_\rho(L)$, in
contrast to that of the variances in the models discussed in
\csect{behavior}, depends, for small $L$, on whether $L$ is even or
odd; this may be thought of as the legacy, at small values of
$\delta$, of the $\delta=0$ behavior of \eqref{v2l}.  Moreover, from
the figure we can observe that:

\begin{itemize}

\item For odd $L$ the variances appear to behave as described in
(P1)--(P5), with $C_1=1/4$, $\lambda_1=0$ and $\lambda_2=3/2$.

\item For even $L$ the ``small'' growth region described by (P2) is
absent: for small and moderate values of $L$ the variances grow as
$C_2(\delta)L^{3/2}$.

\item The values of $\delta$ used in the figure differ by factors of
10, as do the (approximate) straight lines $C_2(\delta)L^{3/2}$ giving
$V(L)$ in the intermediate (and low, for $L$ even) density region.
This suggests that $C_2(\delta)$ vanishes linearly in $\delta$ as
$\delta\searrow0$: $C_2(\delta)=D\delta$ for some constant $D$.  From
\eqref{gammavals} we then have $\gamma_1=2/3$ and $\gamma_2=2$.

 \end{itemize}

The following result, proved in \cite{GLSrigor}, gives a formal
description of this behavior as well as the value of the constant $D$:

\begin{theorem}\label{jpcmmain}Let $N_L$ be the number of particles on
the sites $1,2,\ldots,L$ for the FEP with density $\rho=1/2-\delta$,
  $0<\delta<1/2$.  Then:

 \smallskip\noindent
 (a) For $L$ odd,
 \be\label{thma}
\Var{}_{\nu_\rho}(N_L)\simeq\begin{cases}\frac14,&\text{for $L\ll\delta^{-2/3}$,}\\
  \frac23\sqrt{\frac2\pi}\delta L^{3/2},
          &\text{for $\delta^{-2/3}\ll L\ll\delta^{-2}$,}\\
  \frac14L,&\text{for $L\gg\delta^{-2}$.}\end{cases}
 \ee

 \smallskip\noindent
 (b) For $L$ even,
 \be\label{thmb}
\Var{}_{\nu_\rho}(N_L)\simeq\begin{cases}\frac23\sqrt{\frac2\pi}\delta L^{3/2},
          &\text{for $1\ll L\ll\delta^{-2}$,}\\
  \frac14L,&\text{for $L\gg\delta^{-2}$.}\end{cases}
 \ee
 \end{theorem} Here we are using a slight extension of our earlier
notation: if $A(\delta,L)$ and $B(\delta,L)$ are real valued functions
whose asymptotic behavior in $L$ we wish to compare, and $K(\delta)$
is a positive function, then we write $A(\delta,L)\simeq B(\delta,L)$
for $L\ll K(\delta)$ (respectively for $L\gg K(\delta)$) if for any
$\epsilon>0$ there exists a $\delta_0>0$ and a number $t>0$ such that
$1-\epsilon <A(\delta,L)/B(\delta,L)<1+\epsilon$ when
$\delta<\delta_0$ and $L<tK(\delta)$ (respectively $L>tK(\delta)$).

\begin{remark}\label{more} As mentioned in \crem{gammas}, the $B_i$ of
\eqref{Bvals} have a somewhat different status from the $\gamma_i$ of
\eqref{gammavals}, as we now discuss in the context of the 1d model.
The latter are unambiguously determined by \cthm{jpcmmain}, since
\eqref{thma} and \eqref{thmb} would be false for any $\gamma_1\ne2/3$
and/or $\gamma_2\ne2$ (indeed, even for the 2d model the values of
these exponents are fairly clear directly from the data (see
\cfig{fig:2dscaled})).  But this is not true for $B_1$ and $B_2$, since
\eqref{thma} and \eqref{thmb} are compatible with any choices of
constant coefficients in front of the powers of $\delta$ in the
$L$-ranges of these equations.  No dramatic change in the behavior of
the variance occurs exactly at the $L_i=B_i\delta^{-\gamma_i}$, and we
don't know the size of the transition regions between ``small'' and
``intermediate'' and between ``intermediate'' and ``large''.  The
theorem itself is compatible with these regions being very large
indeed.  Nevertheless, we emphasize that the $B_i$ are sharply
defined, as described in \crem{gammas}, as giving the locations of
the intersections of the asymptotics for the three regions.
\end{remark}

\begin{figure}[ht!]\vskip10pt\begin{center} \hbox to \hsize
{\hss\includegraphics[scale=0.65]{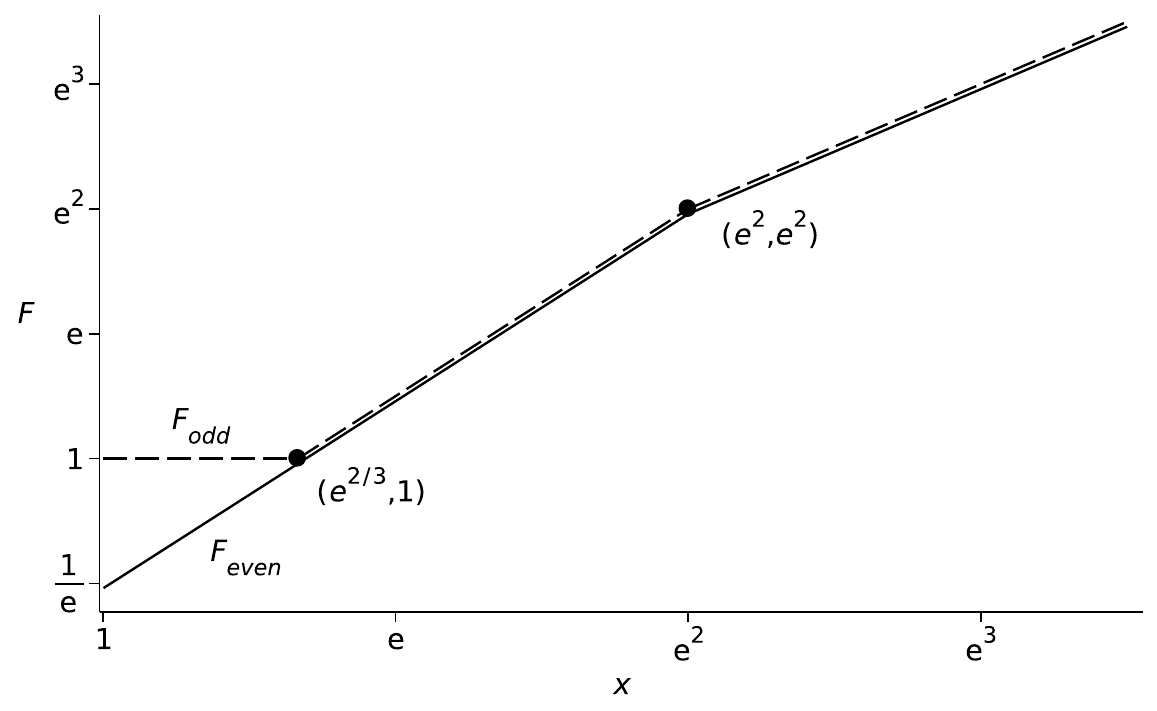}\hss} \caption{Log-log plot
of the limiting functions $F_{\rm odd}$ and $F_{\rm even}$ under power
scaling.  Compare with the inset in \cfig{fig:essence}.}
\label{fig:limits} \end{center}\end{figure}

It is an immediate consequence of \cthm{jpcmmain} that the plots of
$V_\rho(L)$ against $L$ (\cfig{fig:unscaled}) collapse under the power
scaling \eqref{scaling}.  Specifically, if as $\delta\to0$,
$L\to\infty$ through odd (respectively even) values in such a way that
$L^\alpha\to x$ (with $\alpha=\alpha(\delta)=-1/\log\delta$), then
$V_\rho(L)^\alpha\to \Fo(x)$ (respectively
$V_\rho(L)^\alpha\to \Fe(x)$), with
 \begin{align}
\Fe(x)&=\begin{cases}e^{-1}x^{3/2},&\text{for $1\le x\le e^2$,}\\
   x,&\text{for $e^2<x$;}\end{cases}\\
\Fo(x)&=\begin{cases}1,&\text{for $1\le x\le e^{2/3}$,}\\
   \Fe(x),&\text{for $e^{2/3}<x$.}\end{cases}
 \end{align}
 The limiting functions $\Fe$ and $\Fo$ are plotted in
\cfig{fig:limits}.

 \section{Sketch of the proof\label{details}}

In this section we wish to give some hints as to how \cthm{jpcmmain}
can be established, with particular emphasis on a heuristic discussion
of the $V_\rho(L)\simeq D\delta L^{3/2}$ behavior in the intermediate-$L$
(and, for $L$ even, low-$L$) region.  For (setting aside any
difficulties in giving formal proofs) the behavior $V_\rho(L)\simeq1/4$
for $L$ odd and ``small'' (that is, $L\ll\delta^{-2/3}$) is readily
understood in terms of the $\delta=0$ behavior \eqref{v2l} and the
discussion in \csect{behavior}.  The behavior for ``large'' $L$,
$L\gg\delta^{-2}$, is also discussed in \csect{behavior}; on a more
formal level, it is proved in \cite{GLS2} that
 \be\label{fromgls2}
\lim_{L\to\infty}\frac{V_\rho(L)}L=\rho(1-\rho),
 \ee
  which is almost the result of \cthm{jpcmmain}, lacking only the
uniformity as $\delta\searrow0$ with which the limit is achieved.  (The
proof for ``large'' $L$ in \cite{GLSrigor} is a strengthening of the
proof of \eqref{fromgls2} in \cite{GLS2}.)

In the remainder of this section we take $L$ to be a variable
satisfying $1\ll L\ll\delta^{-2}$.  We will freely drop lower-order
terms, approximate sums by integrals, etc.; these manipulations are
justified in \cite{GLSrigor}.

The key ingredient for understanding the $L^{3/2}$ behavior is the
renewal structure of the stationary state $\nu_\rho$, $\rho<1/2$.  In
this state adjacent 1's have probability zero; thus the state is
supported on configurations of the form
 \begin{align}   \label{renew1}
  &\cdots1\ 0\ 1\ 0\ 1\ 0\ 1\ 0\ \hat0\ 1\ 0
   \ 1\ 0\cdots1\ 0\ \hat0\ 1\ 0\ 1\ 0\ 1\ 0
    \cdots1\ 0\ \hat0\cdots\\
  &\hskip50pt=\cdots0\ (1\ 0\ )^{X_{-1}}\ 0\ 
   (1\ 0)^{X_{0}}\ 0\ (1\ 0)^{X_{1}}\ 0
  \ (1\ 0)^{X_{2}}0\cdots.\label{renew2}
 \end{align}
 For $\nu_\rho$ the $00$'s in \eqref{renew1}---or more specifically
the second 0 of such pair, marked as $\hat0$ in \eqref{renew1} and
corresponding to a 0 outside the parentheses in \eqref{renew2}---are
the {\it renewal events} of a {\it renewal process} \cite{GLS2}.  This
means that, if we condition on the occurrence of a such an event at
the origin, so that the $X_i$ may be well defined---for example by
noting that renewal events are separated by distances $2X_i+1$, and
letting $2X_1+1$ be distance from the origin to the first renewal
event on the right---then the $X_i$ are independent random variables
that are identically distributed.

The independence of the $X_i$ is easily understood: because adjacent
empty sites cannot be created during the evolution, the double zeros
in \eqref{renew1} must have been present for all $t$, $0\le t<\infty$,
so that the portions of the system to the left and right of each,
independent under the initial Bernoulli measure, evolve independently.
Note that since under $\nu_\rho$ the probability that a site is
occupied is $\rho$ and the probability of occupied adjacent sites is
zero, the density of the renewal events is $1-2\rho=2\delta$.

 The distribution of the $X_i$ is that of a random variable $X$ with
 \be\label{dist}
P(X=n)=C_n\rho^n(1-\rho)^{n+1}\\
   =\frac{1+2\delta}{2\cdot4^n}C_n(1-4\delta^2)^n,\quad n=0,1,2,\ldots
    \ee
 where $C_n$ is the $n^{\rm th}$ Catalan number,
 \be\label{catalan}
  C_n=\frac1{n+1}\binom{2n}{n}
  \simeq\frac{4^n}{n^{3/2}\sqrt\pi}.
 \ee
 In \eqref{catalan} we have used Stirling's formula. Thus for $n\gg1$
 and $\delta\ll1$,
 \be\label{distasy}
P(X=n)\simeq\frac{1+2\delta}{2n^{3/2}\sqrt\pi}(1-4\delta^2)^n
  \simeq \frac{1}{2n^{3/2}\sqrt\pi}e^{-4\delta^2n}. 
 \ee
 We see that if $\delta$ is zero then $X_i$ has power-law distribution with
a long tail (in particular, with infinite expectation), but that
for positive $\delta$ the distribution is cut off at order $n\approx\delta^{-2}$.
 
Now let $N_L$ and $R_L$ denote respectively the number of particles
and the number of renewal events in the interval $\{1,\ldots,L\}$.  It
is easy to see that
 \be
 N_L=\frac12(L-R_L-\sigma), 
 \ee
 where $\sigma$ may take value $0$, $1$, or $-1$, and thus
 \be\label{NR}
 \Var(N_L)\,\bigl(=V_\rho(L)\bigr)\simeq\frac14\Var(R_L),
 \ee
  up to a correction arising from $\sigma$.  We will show
 (heuristically) below (and prove rigorously in \cite{GLSrigor}) that 
 \be\label{varrl}
\Var(R_L)\simeq\frac83\sqrt{\frac2\pi}\delta L^{3/2} \quad(L\ll\delta^{-2}),
 \ee
 and it is shown in \cite{GLSrigor} that the correction to \eqref{NR}
 arising from $\sigma$ is negligible for $L$ even and is $1/4$ (up to
 order $\delta$) for $L$ odd.  This gives a heuristic justification of
 the $L\ll\delta^{-2}$ cases of \cthm{jpcmmain}.

We now consider \eqref{varrl}.  Let $S_L$ be the number of renewal
events in $\{1,\ldots,L\}$, {\it conditioned on the occurrence of a
renewal event at the origin}. We can express the moments of $R_L$ in
terms of those of $S_L$ by conditioning on the value $Y$ of  the
position of the first renewal event to the right of the origin
($Y\ge1$):
 \be\label{erlk1}
E(R_L^k)=\sum_{y=1}^L P(Y=y) E\bigl((1+S_{L-y})^k\bigr),
   \qquad k=1,2,\ldots,
 \ee
 where $E$ denotes expectation (and we have set $S_0=0$).  Note that
the equality in \eqref{erlk1} does not hold for $k=0$, because the
terms with $y>L$ are omitted from the sum, but that this omission does
not matter for $k\ge1$ because $R_L=0$ if $Y>L$.

 To obtain $P(Y=y)$ we note that, given that there is a renewal event
at $-m$, $m\ge0$, which happens with probability $2\delta$, the
probability that this is the first renewal event at or to the left
of the origin, and that $Y=y$, is $P(X=(m+y-1)/2)$ (and in particular is
0 unless $m$ and $y$ have different parities).  Thus using
\eqref{distasy} we have, for $\delta\ll1$, 
 \begin{align}\nonumber
P(Y=y)&
  =2\delta\sum_{m\ge0,\ m+y\text{ odd}}P\left(X=\frac{m+y-1}2\right)\\
  &\simeq\frac{\delta}{\sqrt\pi}\sum_{n\ge y/2}\frac1{n^{3/2}}
  \simeq\frac{\delta}{\sqrt\pi}\int_{y/2}^\infty\frac{du}{u^{3/2}}
  =2\delta\sqrt{\frac2{\pi y}}. \label{pyey}
 \end{align}
 The moments of $S_L$, the other ingredient in \eqref{erlk1}, may be
obtained from Section~3(ii) of \cite{GodLuck}, once one makes the
approximation that the distance between renewal events is distributed
as $2X$ (rather than $2X+1$), where now $X$ is the random variable
having distribution \eqref{dist} with $\delta=0$.  For it is shown
there that then $S_L\simeq\sqrt{L}|Z|$, with $Z$ a standard normal
random variable, so that 
 \be\label{esls}
E(S_L^2)\simeq L.
 \ee

Now substituting \eqref{pyey} and \eqref{esls} into \eqref{erlk1} we
have, using $1+S_{L-y}\simeq S_{L-y}$,
 \be\label{erl2}
E(R_L^2)\simeq \frac{2\sqrt2\delta}{\sqrt \pi}\int_0^L\frac{L-y}{\sqrt y}\,dy
 = \frac{2\sqrt2\delta L^{3/2}}{\sqrt\pi}\int_0^1\frac{1-u}{\sqrt u}\,du
 = \frac{8}{3}\sqrt{\frac2\pi}\delta L^{3/2}.
 \ee
 Since $L\ll\delta^{-2}$, $E(R_L)^2=(2\delta L)^2\ll E(R_L^2)$, and so
$\Var(R_L)\simeq E(R_L^2)$, yielding \eqref{varrl}.

 It is instructive to note that the probability of at
least one renewal event in $\{1,\ldots,L\}$ is
 \be
P(R_L>0)=\sum_{y=1}^L P(Y=y)
 \simeq2\delta\int_1^L\sqrt{\frac{2}{\pi y}}\,dy
   \simeq4\delta\sqrt{\frac{2}{\pi}}\sqrt L\ll1,
 \ee
 where we have used $L\ll\delta^2$; thus an interval of $L$ sites
typically contains no renewal events.

 \medskip\noindent
 {\bf An unexpected connection:} Within the renewal process framework
just described there is a strong connection between the exponent
$\lambda_2$ in the region of ``intermediate'' $L$ and the coefficient
$C_3$ in the growth relation $V(L)\simeq C_3L$ for ``large'' $L$.
Specifically, the fact that $C_3$ is bounded away from 0 and $\infty$
as $\delta\searrow0$ corresponds, in a sense that we now make precise,
to the value $\lambda_2=3/2$ of the 1d FEP.

Consider then a process of renewal events as described above, but
based on a renewal random variable $X$ with distribution
 \be\label{distasy2}
P(X=n)=\frac{A(\Lambda)}{n^\gamma}e^{-n/\Lambda} ,
  \quad n=1,2,3,\ldots,
 \ee
 with $1<\gamma<2$ and $A(\Lambda)$ a normalization constant (compare
with \eqref{distasy}, which corresponds to $\gamma=3/2$,
$\Lambda=(2\delta)^{-2}$).  For simplicity we assume that $X$ itself,
rather than $2X+1$, gives the distance between renewal events; since
$E(X)\sim\Lambda^{2-\gamma}$ these events have density
$\delta\sim\Lambda^{\gamma-2}$.  We are interested in the behavior of
this process as $\Lambda\nearrow\infty$ (note that $A(\Lambda)$ is
finite and nonzero in this limit).

For $L\ll\Lambda$, which here corresponds to the ``intermediate''
region, we may repeat the calculations above.  Now \eqref{pyey}
becomes $P(Y=y)\sim\delta/y^{\gamma-1}$, and since
$E(S_L^2)\sim L^{2(\gamma-1)}$ from \cite{GodLuck}, following
\eqref{erl2} we find that $E(R_L^2)$ is of order
 \be
\delta\int_0^L\frac{(L-y)^{2(\gamma-1)}}{y^{\gamma-1}}\,dy
 = \delta L^\gamma\int_0^1\frac{(1-u)^{2(\gamma-1)}}{u^{\gamma-1}}\,du,
 \ee
 i.e., of order $\delta L^\gamma$. Since
$E(R_L)^2=(\delta L)^2\ll\delta L^{\gamma}$, $\Var(R_L)$ is of this
same order; thus $\lambda_2=\gamma$.
  
The asymptotic value of $\Var(S_L)$ is obtained in \cite{Feller}
(see Problems 19--23 of Chapter XIII) and the same methods may be used
to determine $\Var(R_L)$:
 \be\label{varrl2}
\Var(R_L)\simeq\frac{\Var(X)}{E(X)^3}\,L.
 \ee
  (In fact, \eqref{varrl2} also holds with $\Var(R_L)$ replaced by
$\Var(S_L)$).  We observed above that $E(X)\sim\Lambda^{2-\gamma}$,
and a similar calculation shows that $E(X^2)\sim\Lambda^{3-\gamma}$
and so also $\Var(X)\sim\Lambda^{3-\gamma}$, so that from
\eqref{varrl2}, $\Var(R_L)\sim\Lambda^{2\gamma-3}L$.  Thus the
coefficient of $L$ in the region of ``large'' $L$ is of order unity as
$\Lambda\nearrow\infty$ precisely when the ``intermediate'' region
exponent $\lambda_2=\gamma$ has value $3/2$.

\section{Related models and open problems\label{conclusion}}

As indicated in \csect{model}, the results that we describe there and
in \csect{details} apply to several other one-dimensional models.  The
continuous-time model that we treat in \csectt{behavior}{details}, in
which particles jump to all directions at equal rates, may be
generalized in one dimension to the {\it partially asymmetric} model,
in which left and right jumps occur at different rates.  These models
again have an absorbing phase transition at $\rho_c=1/2$, and it is
shown in \cite{agls} that the limiting measure $\nu_\rho$ for
$\rho<1/2$ (and indeed for all $\rho$) is independent of the degree of
asymmetry, so that for $\rho<1/2$ it is the measure analyzed in
\csect{details}.  The latter result is also true in discrete time for
the {\it totally} asymmetric FEP \cite{GLS2}, in which particles jump
only to the right and then, of course, only if their left-hand
neighboring site is occupied (although here the $\rho>1/2$ measures do
not coincide with those of the continuous-time model).  Thus
\cthm{jpcmmain} holds for all these models.

Given the above, we can understand the structure of the measure
$\nu_\rho$, $\rho<1/2$, described in \csect{details}, by viewing it as
the limiting measure for the totally asymmetric model.  To do so we
associate with each initial configuration of that model a random walk
on the integers, in which the walker takes a step to the right
(respectively left) at time $i$ if and only if site $i$ is empty
(respectively occupied).  Since $\rho<1/2$ the walk has a drift to the
right.  It is then easy to convince oneself that the final
configuration is completely determined by the initial one, and that
the renewal events in \eqref{renew1} occur at sites $i$ such that the
random walk reaches a new maximum at time $i$.  The intervals between
two such maxima, which have the form $2X_i+1$ in \eqref{renew2}, are
certainly independent, and their distribution \eqref{dist} is
determined by the well-known fact that the Catalan number $C_n$ counts
the number of walks ({\it Dyck paths of length $2n$}) between two such
maxima which are separated by a distance $2n+1$.

The situation is different for the {\it symmetric} discrete-time model
in one dimension \cite{GLS3} (partially asymmetric discrete-time
models have not been studied, to our knowledge).  In contrast to the
models discussed in the previous paragraph, adjacent empty sites can
be created during the evolution, and the measure $\nu_\rho$ of
\csect{details} is no longer the limiting measure.  In particular, at
$\rho=1/2$, where the measure \eqref{muhalf} is the unique translation
invariant stationary (TIS) measure for the models of the previous
paragraph, now there are two uncountable families of extremal TIS
measures, each in correspondence with the set of all TI measures on
$\{0,1\}^\bbz$; these describe situations in which patterns formed
by occupied and unoccupied sites move to the left or right,
respectively, with velocity 2.  It is shown in \cite{GLS3} that
$\lim_{t\to\infty}\mu_t^{(\rho)}$ exists for $\rho<1/2$, but little is
known about the nature of the limiting measure.  For $\rho=1/2$ the
existence of the limit is not proven; if it exists it must of course
be a convex combination of the extremal TIS measures mentioned above,
and simulations suggest a highly nontrivial combination.

Hexner and Levine \cite{hl} study several models, other than the FEP,
which exhibit a phase transition to an absorbing state: the Manna
model in one dimension and the ``random organization'' model in both
one and two dimensions.  In all cases they find that the state at the
critical value of the density is hyperuniform.  Their plots showing
the dependence of the fluctuations on $L$ and on the density are
generally similar to our \cfig{fig:essence}, so we suspect that the
phenomenology that we describe in \csect{behavior} will apply there,
also, but we have not investigated the question.

The complete absence of theoretical results for the FEP models in
dimensions two and higher presents a challenge to mathematical
physicists.  For example, one would very much like to have a proof of
the existence of active states at some density below $1/2$ and/or,
conversely, a proof that the system must freeze when started from a
Bernoulli initial distribution of sufficiently low density $\rho$
(say, $\rho<10^{-23}$).  On a more refined level one would like to
establish the existence of a critical density $\rho_c<1/2$ such that
final states are frozen for $\rho<\rho_c$ and active for
$\rho>\rho_c$, and then prove that, as suggested by simulations, the
final state at $\rho_c$ is frozen and hyperuniform.


\begin{thebibliography}{99}

\bibitem{agls}A. Ayyer, S. Goldstein, J. L. Lebowitz, and E. R. Speer,
Stationary States of the One-dimensional Facilitated Asymmetric Exclusion
Process. arXiv:2010.07257 [math.Pr]. To appear in AIHP.

\bibitem{BESS} Oriane Blondel, Cl\'ement Erignoux, Makiko Sasada, and
Marielle Simon, Hydrodynamic Limit for a Facilitated Exclusion Process.
{\it Annales de l'Institut Henri Poincar\'e, Probabilit\'es et
Statistiques} {\bf 56}, 667714 (2020).

\bibitem{BES} Oriane Blondel, Cl\'ement Erignoux, and Marielle Simon,
Stefan Problem for a Non-Ergodic Facilitated Exclusion Process.  {\it
Probability and Mathematical Physics}
{\bf 2}, 127--178 (2021).

\bibitem{del}B. \~Derrida, C.~Enaud, and J.~L.~Lebowitz, The
  Asymmetric Exclusion Process and Brownian Excursions.  {\it
    J.~Stat.~Phys.} {\bf 115}, 365--382 (2004). 

\bibitem{Feller} W. Feller, {\it An Introduction to Probability Theory
and its Applications, Volume I}, Third edition.  John Wiley \& Sons,
New York, 1968.

\bibitem{GodLuck} C.~Godr\`eche and J.~M.~Luck, Statistics of
  Occupation Time of Renewal Processes.  J. Stat. Phys. {\bf 104},
  489--524 (2001).

\bibitem{glsshort}S. Goldstein, J. L. Lebowitz, and E. R. Speer, Exact
Solution of the F-TASEP.  J. Stat. Mech. 123202 (2019).

\bibitem{GLS2} S. Goldstein, J. L. Lebowitz and E. R. Speer, The
Discrete-Time Facilitated Totally Asymmetric Simple Exclusion Process.
Pure Appl. Funct. Anal. 6, 177203 (2021), arXiv:2003.04995 [math-ph].

\bibitem{GLS3} S. Goldstein, J. L. Lebowitz and E. R. Speer, Stationary
States of the One-Dimensional Discrete-Time Facilitated Symmetric
Exclusion Process.   arXiv:2201.05175 [math.Pr]. 

\bibitem{GLSrigor} S. Goldstein, J. L. Lebowitz and E. R. Speer,
Approach to Hyperuniformity in a Simple 1-d system.  In preparation.

\bibitem{GL}Ghosh, Subhroshekhar, and Lebowitz, Joel L., Fluctuations,
Large Deviations and Rigidity in Hyperuniform Systems: A Brief Survey.
{\it Indian Journal of Pure and Applied Mathematics} {\bf48 (4)} 609--631
(2017).

\bibitem{hl}Daniel Hexner and Dov Levine,  Hyperuniformity of
Critical Absorbing States. {\it Phys. Rev. Lett.} {\bf 114}, 110602
(2015).

\bibitem{Liggett}T. M. Liggett, {\it Interacting Particle Systems.}
Springer-Verlag, New York, 1985.

\bibitem{lub}S. L\"ubeck, Scaling Behavior of the Absorbing
  Phase Transition in a Conserved Lattice Gas Around the Upper
  Critical Dimension.  {\it Phys. Rev. E} {\bf 64}, 016123 (2001).
 
 \bibitem{oliveira}M\'ario J. Oliveira,  Conserved Lattice Gas Model with
   Infinitely Many Absorbing States in One Dimension.  {\it Phys. Rev. E} 
   {\bf 71}, 016112 (2005).

\bibitem{ps} E. Presutti and H. Spohn, Hydrodynamics of the Voter
  Model.  {\it Ann. Probab.} {\bf 11}, 867--875 (1983).

\bibitem{rpv}Michela Rossi, Romualdo Pastor-Satorras, and Alessandro
Vespignani, Universality Class of Absorbing Phase Transitions with a
Conserved Field. {\it Phys. Rev.  Lett.} {\bf85}, 1803 (2000).

\bibitem{T}Torquato, Salvatore, Hyperuniform States of Matter.  {\it
Phys. Rep.} {\bf 745}, 1--95 (2018).

\end{thebibliography}
 \end{document}